\documentclass[prl,aps,twocolumn]{revtex4}
\usepackage{graphicx}
\begin{document}

\newcommand\SE{\mathrm{SE}}%
\newcommand\short{\mathrm{short}}%
\begin{flushleft}
\textbf{Comment on ``Josephson Current through a Nanoscale Magnetic
  Quantum Dot''}
\end{flushleft}

In a recent work~\cite{Siano04a}, Siano and Egger (SE) studied the Josephson
current through a quantum dot in the Kondo regime using the quantum
Monte Carlo (QMC) method.
Several of their results were unusual, and inconsistent
with those from the numerical renormalization group (NRG)
calculations\cite{ChoiMS04c,Yoshioka00a} among other previous studies.
Those results in Ref.\cite{Siano04a} are not reliable for the following
two reasons: (i) The definition of the Kondo temperature was wrong; (ii)
There were substantial finite-temperature effects.

We first clarify point (i).
The \emph{normal-state} Kondo temperature\cite{Haldane78a,HewsonBook}
\emph{in the absence of superconductivity} provides one of the most
significant energy scales of the system.
SE defined the Kondo temperature as
\begin{equation}
\label{comment::eq:TK:SE}
T_K^\SE=
\exp[\pi\epsilon_0(\epsilon_0+U)/\Gamma_\SE U]
\sqrt{\Gamma_\SE U}/2
\end{equation}
with $\Gamma_\SE=2\pi\rho_0|t|^2$, where $|t|^2$ denotes the coupling to
\textit{one lead} and $\rho_0$ the density of states (DOS) at the Fermi
level.
In Ref.\cite{ChoiMS04c} we 
defined it as
\begin{equation}
\label{comment::eq:TK}
T_K=
\exp[\pi\epsilon_0(\epsilon_0+U)/2\Gamma U]
\sqrt{\Gamma U/2} 
\end{equation}
with $\Gamma=2\pi N_0|V|^2$, where $|V|^2$ denotes the coupling to
\textit{one lead} and $N_0$ the DOS at the Fermi level \emph{per spin}
[the factor $2$ in the coupling comes from the two leads].
It is important to clarify the difference between the two definitions
since different definitions of $T_K$ result in significantly different
scaling behaviors of physical quantities.
%
We note that both forms,
Eqs.~(\ref{comment::eq:TK:SE}) and (\ref{comment::eq:TK}), appear in the
literature.  However, in Eq.~(\ref{comment::eq:TK:SE}) $\Gamma_\SE$
should be the \emph{full} width at half maximum of the single particle
level of the noninteracting dot~\cite{vanderWiel00a}, whereas in
Eq.~(\ref{comment::eq:TK}) $\Gamma$ should be the \emph{half} width at
half maximum (HWHM) of the single particle level.
To see the precise meaning of $\Gamma_\SE$, let us take the limit
$\Delta=0$ and $U=0$ in the local Green's function (GF) in Eq.~(6) in
Ref.~\cite{Siano04a}.  Going over to the retarded GF, we find
$G_R^{-1}\sim E+i\Gamma_\SE$, which yields the spectral function
\begin{equation}
A(E) = -\frac{1}{\pi}\mathrm{Im}G_R
=\frac{1}{\pi} \frac{\Gamma_\SE}{E^2+\Gamma_\SE^2}\,.
\end{equation}
Therefore, $\Gamma_\SE$ corresponds to \emph{HWHM}.  Namely, the two
hybridization $\Gamma_\SE$ and $\Gamma$ are the same. Therefore, the two
Kondo temperatures in Eqs.~(\ref{comment::eq:TK:SE}) and
(\ref{comment::eq:TK}) are related with each other by
\begin{math}
T_K^\SE = T_K^2 / \sqrt{\Gamma U}\,,
\end{math}
which implies that the scale $\Delta/T_K^{\SE}$ differs from the scale
given in our work \cite{ChoiMS04c}.
The unusual definition of Kondo temperature in
Eq.~(\ref{comment::eq:TK:SE}) explains the (otherwise) unusual behaviors
of $I(\phi)$ with respect to $U/\Delta$ in Fig.~2 of SE.

\begin{figure}[thbp]
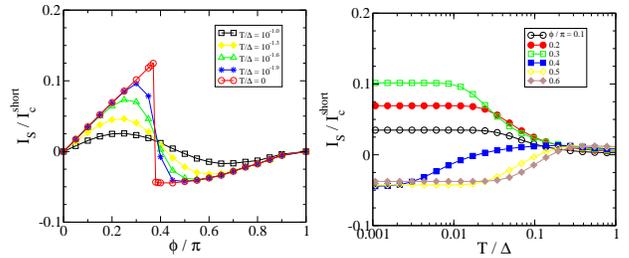

\centering
\includegraphics*[width=40mm]{fig1a}
\includegraphics*[width=40mm]{fig1b}
\caption{(a) Josephson current $I_S(\phi)$ at different temperatures.
  (b) Josephson current as a function of temperature for different
  values of $\phi$. $\Delta/T_K=1.6$.}
\label{comment::fig:1}
\end{figure}

We now move on to point (ii).  In Ref.~\cite{Siano04a} all calculations
have been done at a finite temperature $T=0.1\Delta$ and SE note that
``this appears to be quite close to the ground-state limit''.
This is particularly important in the determination of the current-phase
relation. To estimate the Josephson energy we note that it is obtained
from
\begin{equation}
E_J(\phi) = \int^\phi{d\phi'}\; I_S(\phi')
\sim \Delta \frac{I_c}{I_c^\short} \,,
\end{equation}
where $I_c$ is the effective critical current of the system and
$I_c^\short\equiv{}e\Delta/\hbar$ the critical current of the open
contact.
According the the numerical results in Ref.~\cite{Siano04a},
$I_c/I_c^\short\leq 0.1$ for $\Delta/T_K^\SE\gtrsim 5$
($\Delta/T_K\gtrsim 1$ in Ref.~\cite{ChoiMS04c}). We think that in
most plots in Ref.~\cite{Siano04a} the current-phase relation contains
significant amount of thermal activation.  
To confirm this we have performed
NRG calculations at finite temperatures and the results in
Fig.~\ref{comment::fig:1} demonstrate the strong finite-temperature effects.
The sharp transition at zero temperature is washed out and the critical
current is reduced by a factor of 5 for $T/\Delta=0.1$.
The discrepancy between the NRG and QMC data in the new Fig.~2 of the
Reply\cite{Siano04b} may simply reflect the different estimates of
critical value $\Delta_c/T_K$ (i.e., the NRG and QMC data are in
different phases), and may not be an evidence that the NRG is less
accurate.

\begin{flushleft}
\textbf{Authors:}\\
Mahn-Soo Choi and Minchul Lee,
Department of Physics, Korea University, Seoul 136-701, Korea\\
Kicheon Kang, Department of Physics, Chonnam National University,
Gwang-ju
500-757, Korea\\
W. Belzig, University of Basel, Klingelbergstrasse 82, 4056 Basel,
Switzerland
\end{flushleft}


\end{document}